\documentclass{aastex}
\usepackage{spr-astr-addons}
\usepackage{url}
\usepackage[bookmarks,colorlinks,citecolor=cyan,breaklinks]{hyperref}

\RequirePackage{color}

\def\teff{$T\rm_{eff }$}
\def\hst{{\it HST\/}}

\def\ngc#1{\hbox{NGC\,#1}}
\def\feh{\hbox{\rm [Fe/H]}}
\def\afe{\hbox{\rm [$\alpha$/Fe]}}

\begin{document}

\title{WFPC2 UV survey of Galactic Globular Clusters. The Horizontal Branch
temperature distribution.}

\shorttitle{WFPC2 UV survey: HB temperature distributions of GGCs}
\shortauthors{Lagioia E.~P. et al.}

\author{Lagioia E.~P.} 
\author{Dalessandro E.} 
\author{Ferraro F.~R.} 
\author{Lanzoni B.} 
\affil{Department of Physics and Astronomy, University of Bologna, Italy}
\author{Salaris M.} 
\affil{Astrophysics Research Institute, Liverpool John Moore University, UK}
\email{edoardo.lagioia2@unibo.it}


\begin{abstract}
Ultraviolet observations are the best tool to study hot stellar populations
which emit most of their light at short wavelengths. As part of a large
project devoted to the characterization of the UV properties of Galactic globular
clusters (GGCs), we collected mid/far UV and optical images with the WFPC2@\hst\
for 31 GGCs. These clusters cover a wide range in metallicity and structural
parameters, thus representing an ideal sample for comparison with theoretical
models. Here we present the first results from an ongoing analysis aimed at
deriving the temperature distribution of Horizontal Branch stars in GGCs.
\end{abstract}


\section{Introduction}\label{sec:intro}
In old stellar populations like Galactic globular clusters (GGCs), the hottest
stars are Blue Stragglers (BSS), Horizontal Branch (HB), Post-Early Asymptotic
Giant Branch (PEAGB) and AGB-manqu\'e whose spectral emission peaks at effective
temperatures higher than $\sim 6,000\,K$ \citep{Fer12,Moe07,Fer97,Swe87}. Therefore
these stars mostly emit in the ultraviolet (UV) part of the electromagnetic
spectrum \citep{Moe95,deB85}. Since the Earth's atmosphere is almost completely
opaque to this wavelength regime, any campaign for the observation of hot stars
in UV must be performed using space facilities like, for instance, the Hubble
Space Telescope (\hst) or the Galaxy Evolution Explorer (GALEX). Unfortunately,
the theoretical models describing the evolution of stars of the earliest
spectral types still suffer from many uncertainties which are also due to the
absence of a complete and homogeneous sample of UV emitters. Therefore, the UV
observation of hot stars in GGCs is extremely important in order to compare data
and evolutionary models in observational planes not hampered by critical color -
effective temperature (\teff) transformations
\citep{Dal13,Sch12,Dal11,Fer06-2,Cas04,Fer03,Fer98,Fer97-2}. 

One of the most interesting stellar evolutionary sequences that still wait to be
fully understood is the HB. It is commonly accepted that metallicity is the
first parameter driving the Horizontal Branch morphology. Indeed, metal-rich
GGCs typically have red HBs while metal-poor ones have more extended and blue
HBs. However, there are several clusters with the same metallicity showing
remarkable differences in Horizontal Branch morphology: nice examples are the
couples \ngc6388 - \ngc5927, M\,3 - M\,13 and M\,15 - M\,92. Therefore,
metallicity alone is not able to explain the observation of the complex HB
zoology in GGCs \citep{Fre81}. This issue, known as the `2nd parameter' problem,
has focused the attention of several authors in the last years
\citep{Gra10,Dot10,Cat09,Lee94,Fus93}, but the solution is not obvious because
many mechanisms play a role in shaping the color distribution of HB stars, such
as mass loss, age and He abundance \citep{Roo73}. Nonetheless, now there is a
general consensus about the fact that age is the main global 2nd parameter
\citep{Gra10,Dot10}.

\par The HB temperature extension can be an efficient way to parametrize the HB
morphology and therefore to study the `2nd parameter' problem. This approach has
been already attempted by \citet{Rec06} and, later on, by \cite{Bab09}, who used
optical CMDs for their analysis. However, temperature derived from optical
filters only, can be underestimated by more than $10,000\,K$ in the case of very
extended HBs \citep[see the case of \ngc2808, ][]{Dal11}. This demonstrates the
difficulties of deriving properties of hot HB populations from pure optical
data. Indeed in the optical plane, stars move down to the blue HB tail either
because the bolometric correction increases or because stellar luminosity
decreases. For very blue HB stars (like Blue Hook stars) both the above
quantities are changing and optical colors cannot provide a reliable estimate of
\teff.     

\par As a part of a large project aimed at characterizing the UV properties of
GGCs \citep{San14,Dal13-2,Dal13,Con12,Dal12,San12,Sch12,Dal11,Dal09} we present
here some preliminary results coming from the photometric analysis of UV
observations of 31 GGCs obtained with the Wide Field and Planetary Camera 2
(WFPC2) and finalized to accurately derive the HB temperature distribution
and extension.

\section{Observations, reduction and data analysis}\label{sec:observ} 
A large dataset~\footnote{Program GO 11975, PI F.~R. Ferraro.} of mid/far UV and
optical images, in the filters $F160BW, F170W, F255W, F300W, F336W$ and $F555W$,
has been collected with the WFPC2 on-board \hst and has been complemented with a
few images collected in the $F150LP$ filter with the Solar Blind Channel (SBC)
of the Advanced Camera for Survey (ACS@\hst), for a total sample of 31 GGCs.
The targeted GGCs cover a wide range of metallicity ($-2.2<\feh<-0.4$) and
concentration parameter ($0.8<\rm{c}<2.5$). 

Every WFPC2 frame was pre-processed through the standard \hst/WFPC2 pipeline for
bias subtraction, dark correction and flat fielding and corrected for
specific instrument-induced variation of the signal \citep[`34th row' effect and
Pixel Area Correction, ][]{Bag02} across the field of view (FoV).  Then, we have
extracted single images from the four-chip mosaic of the WFPC2 and, for each given
chip and filter, we have combined them by using specific IRAF tasks and cosmic-ray
rejection algorithms~\footnote{Cosmic ray contamination is particularly
important for long-exposed UV images.}, in order to obtain a combined median
image with high S/N. 
  
The photometric analysis of the combined image has been performed with
DAOPHOT\,IV/ALLSTAR \citep{Ste87}. An analytic point spread function (PSF) has
been evaluated on each image by selecting bright, isolated stars uniformly
distributed over the entire chip area. Once the instrumental catalogs have been
obtained for the different chips and filters, they have been geometrically
matched by using DAOMATCH/DAOMASTER \citep{Ste94}, in order to build a master
catalog. Stars in the master catalog are then force-fitted onto single images by
using ALLFRAME \citep{Ste94}, eventually obtaining an instrumental catalog for
each filter/chip combination. For every catalog we computed and applied the
aperture correction by using DAOGROW \citep{Ste90} and the Charge Transfer
Efficiency (CTE) and UV loss by means of the equations provided by \cite{Dol09}.
Instrumental magnitudes were finally calibrated to the VEGAMAG system by
following the prescriptions and the zero points provided in the WFPC2 Instrument
Science Report 97-01 \citep{Bag97}.  

\section{UV-optical CMDs: the case of $\omega\,Cen$}\label{sec:cmds} 
Among the surveyed targets particular interest is undoubtedly reserved to
$\omega\,Centauri$ \citep[\ngc5139, ][]{Sol05,Bed04,Fer04}. This stellar system
hosts at least six sub-population of stars with different $\alpha$-elements and
iron abundances \citep{Pan11,Gra11,Bel10,Cal09,Joh09,Sol07,Sol04,Ori03}, and with
significant spread in helium \citep{Pio05}.  As far as the hot stellar
populations are concerned, $\omega\,Cen$ shows, together with \ngc2419
\citep{Dal08}, the largest population of BSSs \citep{Fer06}, as well as one of
the most extended HB \citep{Dcr00}.  

Due to its large half-light radius of 5\,arcmin \citep[][2010 edition]{Har96}, 13
fields around the center of $\omega\,Cen$ have been imaged in the WFPC2 UV
survey. Images were collected in two filters: $F170W$ and $F555W$.   

\begin{figure}
\centering
\includegraphics[width=\columnwidth]{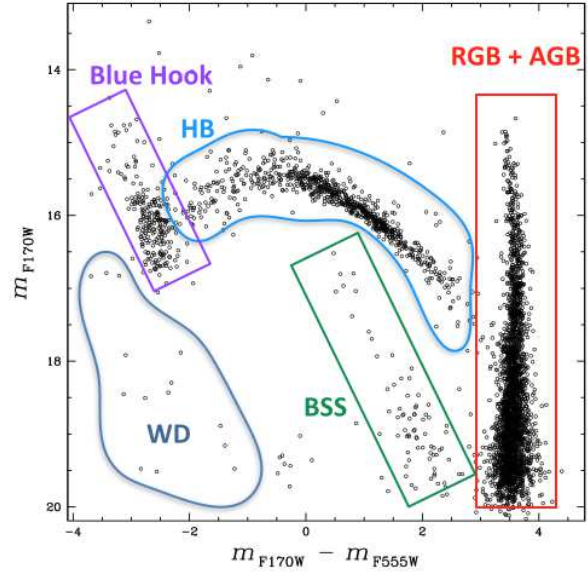}
\caption{(${\rm m}_{F170W}, {\rm m}_{F170W}-{\rm m}_{F555W}$) CMD of
$\omega\,Cen$. Encircled areas emphasize the different evolutionary sequences:
RGB+AGB = Red Giant Branch + Asymptotic Giant Branch, BSS = Blue Stragglers, HB
= Horizontal Branch, Blue Hook, WD = White dwarfs.} 
\label{fig:fig1} 
\end{figure}

The final CMD of the total FoV of $\omega\,Cen$ is shown in Fig.~\ref{fig:fig1}.
In this plane we can distinguish at least three well defined sequences. The
vertical strip at (${\rm m}_{F170W}-{\rm m}_{F555W}$) $\approx 3.5$ is a
combination of the bright Red Giant Branch (RGB) and AGB stars. The presence of
such cool stars in this UV plane is due to the spectral response of the filter
F170W which suffers from a quite important `red-leak' (WFPC2 Instrument
Handbook, August 2008).  The diagonal strip extending from the base of the RGB
to ${\rm m}_{F170W} \sim 16.5$ is populated by the brightest BSSs in our
FoV~\footnote{We have identified the BSSs by matching our catalog with the
candidate BSSs selected by \cite{Fer06}.}. The brightest and hottest stars in
the (${\rm m}_{F170W}, {\rm m}_{F170W}-{\rm m}_{F555W}$) CMD are HB and post-HB
stars. The HB of $\omega\,Cen$ is almost homogeneously populated from red (${\rm
m}_{F170W}-{\rm m}_{F555W} \lesssim 3$) to extremely blue colors (${\rm
m}_{F170W}-{\rm m}_{F555W} \approx -2$).  Fig.~\ref{fig:fig1} shows that at the hot
end of the HB there is an uprising sequence populated by Blue Hook stars
\citep[${\rm m}_{F170W} \lesssim 17, ({\rm m}_{F170W}-{\rm m}_{F555W}) \gtrsim
-2$, ][]{Dan10,Cas09}.

\begin{figure}
\centering
\includegraphics[width=\columnwidth]{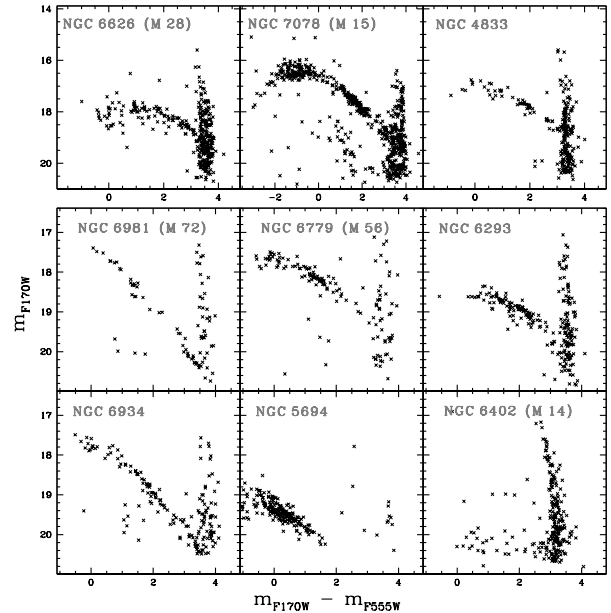}
\caption{(${\rm m}_{F170W}, {\rm m}_{F170W}-{\rm m}_{F555W}$) CMD of nine out of
31 GGCs present in out survey.} 
\label{fig:fig2}
\end{figure}

Other than for $\omega\,Cen$, the data analysis has been completed for nine
additional GGCs in the survey. The relative CMDs are shown in
Fig.~\ref{fig:fig2}. They display features (RGB, HB, BSS) similar to those
already described for the CMD of $\omega\,Cen$. Cluster-to-cluster differences
of course may appear, in particular each cluster shows a quite different HB
extension (Lagioia et al.  2014, in preparation). 

\section{The HB temperature distribution of \ngc6293}
\begin{figure}
\centering
\includegraphics[width=\columnwidth]{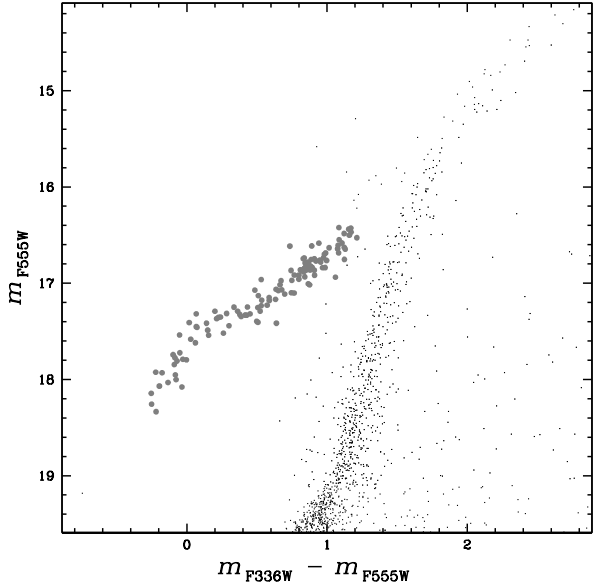}
\caption{Optical (${\rm m}_{F555W}, {\rm m}_{F336W}-{\rm m}_{F555W}$) CMD of
\ngc6293. The filled gray circles mark the HB stars selected in this plane.} 
\label{fig:fig3}
\end{figure}

As stated in Sect.~\ref{sec:intro}, one of the scientific goals of the WFPC2 UV
survey is to derive the temperature distribution of HB stars in the observed
clusters. In order to accomplish this task we must compare the observations with
suitable theoretical models. As detailed in \cite{Dal13,Dal11} and \cite{Sal13},
there are several important points that should be properly taken into account:
the effect of radiative levitation, $\alpha$-element variation and the presence
of sub-populations with different He abundance. In this sense $\omega\,Cen$
represents a very interesting case but it needs particular caution because of
its complexity. Among the clusters already available (see Figures~\ref{fig:fig1}
and \ref{fig:fig2}) we decided to start our analysis with a relatively simple
case: \ngc6293. This GGC has apparent distance modulus $(m-M)_{\rm V} = 16.00$
and reddening $E(B-V) = 0.36$ \citep[][2010 edition]{Har96} and it is the most
metal-poor cluster observed in the Galactic bulge \citep[$\feh = -1.73$,
][]{Val07}. The optical CMD of this cluster is shown in Fig.~\ref{fig:fig3}. 

\begin{figure}
\centering
\includegraphics[width=\columnwidth]{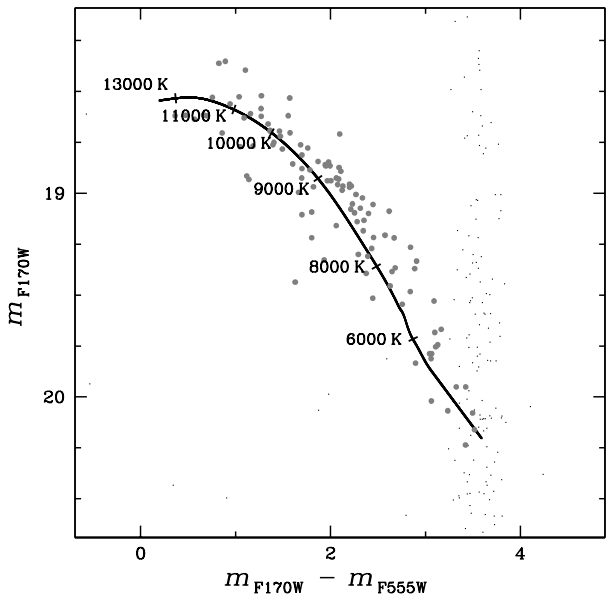}
\caption{Far UV - optical ($\rm m{}_{F170W}, {\rm m}_{F170W}-{\rm m}_{F555W}$)
CMD of \ngc6293. The filled gray circles correspond to the HB stars selected in the
optical CMD (Fig.~\ref{fig:fig3}). A ZAHB model, suitably calculated for this
cluster (see text for explanation), has been overplotted as a solid black line.
Some temperature values along the ZAHB are also marked.} 
\label{fig:fig4}
\end{figure}

\begin{figure}[t]
\centering
\includegraphics[width=\columnwidth]{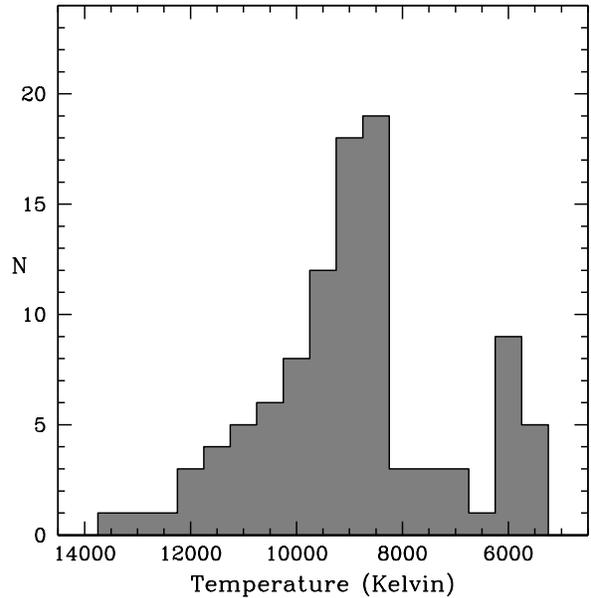}
\caption{Histogram of the temperature distribution of HB stars of \ngc6293
obtained from the analysis of the (${\rm m}_{F170W}, {\rm m}_{F170W}-{\rm
m}_{F555W}$) plane.} 
\label{fig:fig5}
\end{figure}

Because of the `red-leak' of the $F170W$ filter (Sect.~\ref{sec:cmds}) some of
the coolest HB stars may overlap the RGB sequence in the far UV - optical plane.
For this reason we first selected HB stars in the optical CMD (filled gray
circles in Fig.~\ref{fig:fig3}), where the separation between HB and RGB stars
at low \teff\ is clearer. 
The position of the HB stars in the UV-optical CMD is
shown in Fig.~\ref{fig:fig4}, where we also plotted the best-fit Zero Age
Horizontal Branch (ZAHB) model from the Basti Database \citep{Pie06}, computed
by assuming $\feh = -1.84$, $\afe = 0.4$, $(m-M)_0 = 15.2$ and reddening $E(B-V)
= 0.36$. It is important to notice that, because of the spectral response of the
$F170W$ filter, the reddening correction is a strongly variable function of the
temperature (color) of the star and it therefore requires particular attention.
A crucial step in our analysis is the search for variable stars populating
the HB. Indeed, they could considerably affect the determination of the HB
temperature distribution. According to the Catalogue of Variable Stars in
Galactic Globular Clusters \citep[][2013 update]{Cle01}, \ngc6293 is known to
host five RR Lyrae stars but only three of them fall in our WFPC2 FoV.
Moreover, their position in the Clement's catalog is derived from ground-based
observations and we were not able to unambiguously identify them in our
high-resolution HST catalog. Their inclusion in our HB sample, however, does not
affect at all our analysis.

Following the procedure described in \cite{Dal11} we derived the temperature of
each star by projecting its color onto the ZAHB model. The temperature is the
result of the interpolation between the two closest points of the ZAHB. The
distribution obtained is shown in Fig.~\ref{fig:fig5}.

We observe that the histogram spans a temperature interval going from $\sim
5000\,K$ to $\sim 13000\,K$ and that the mode of the distribution is located
at $\sim~8500\,K$. More than 50\% of the stars have a temperature higher than
the peak value. To establish how much the `red-leak' affects the estimate of
temperature of the cold HB stars in the UV-optical plane, we determined the
temperature of the same stars also from the optical ($m_{F255W}-m_{F555W},
m_{F255W}$) CMD. Then, we determined the difference between the two temperature
estimates. We found that the average differences of the HB stars colder and hotter
than $\sim 8000\,K$ agree each other within the errors. 

Unfortunately, neither photometric nor spectroscopic temperature estimates for
the HB stars of \ngc6293 are available in literature. Therefore, no direct
comparison can be done with our data.  Nevertheless, \cite{Dal11}, by using the
same set of theoretical models described in the present paper, found that the
photometric estimates of the HB star temperatures of the cluster \ngc2808 are
fairly consistent with the temperatures derived from the spectroscopic
observations by \citep{Moe04}.

\section{Summary}
We have analyzed far/mid UV-optical images for ten out of 31 GGCs collected
within the WFPC2 UV survey. The main task of this analysis is to derive the
extension and the temperature distribution of the HB of the targeted clusters
taking advantage of the UV data. The use of UV bands is the best tool to
accomplish our task since, for stars with high \teff, the bolometric correction
decreases and the stellar luminosity increases with respect to the optical
bands. It has been shown, indeed, that the temperatures of extended HBs derived
from a combination of pure optical filters can be underestimated by $10,000 -
15,000\,K$ \citep{Dal11}. We used \ngc6293 as a starting point in our analysis,
determining the temperature distribution of its HB stars. We are now applying
this analysis to other GGCs in our sample with extended HBs, like $\omega\,Cen$
and M\,15.

\acknowledgments
This research is part of the project COSMIC-LAB funded by the
European Research Council, under contract ERC-2010-AdG-267675.

\bibliographystyle{spr-mp-nameyear-cnd}

\end{document}